\documentclass[]{cupconf}
\usepackage{graphicx}

\title[Starbursts and their contribution to metal enrichment]
      {Starbursts and their contribution \\ to 
               metal enrichment}
  
\author[Daniel Kunth]{\\ Daniel Kunth$^1$}   
\affiliation{$^1$ Institut d'Astrophysique  de Paris, F-75014 Paris, France; kunth@iap.fr}   

\begin{document}
\maketitle

\begin{abstract}
I review the properties of Starburst galaxies, compare the properties of the local ones with more distant starburts and
examine their role in the metal enrichment of the interstellar medium and the intergalactic-intracluster medium. Metallicity is not an arrow of time and contrary to current belief metal rich galaxies can also be found at high redshift.
\end{abstract}

\firstsection

\section{Introduction}

 For the purpose of this conference, let me first stress  that known starburst galaxies  in the local universe are not metal rich (Z$\leq Z_{\odot}$). Having written this, I remark however that 'metal-rich' has different acceptance depending upon one's field of interest. Stellar astrophysicists deal with individual  stars as metal poor as [X] $\approx -4$ similar or quite to some DLAs, hence they would regard solar abundances as being large! On the other hand, the most dramatic case known, so far, of metal overabundance comes from QSOs spectroscopy with [N/H] $\approx 1$ (Hamann et al. 2002). Most of my talk will review our knowledge about local starbursts. While starburst galaxies are not known to be  metal rich objects they fit nethertheless with this meeting. They are particularly interesting in their own right, while bearing very strong similarities with their high redshift counterparts, because they are very significant components of our present-day universe and mostly important sources of chemical enrichement. Moreover, their study, brings the important clue that metallicity is not providing us with an arrow of time, as is too often believed. Moreover, high redshift galaxies that are obscured and with high star formation rates can be even more metal rich than their low luminosity analogs of the Local Universe.

\section{Phenomenology and Definitions}

The concept of "starburst" sometimes leads to confusion. A starburst "event" refers to the formation of thousands of massive stars ($M_{\odot} \geq 20$) on a very small volume (few pc) and a time scale of only a few million years (Hoyos 2006). Such an event may occur in a normal galaxy such as our Galaxy or 30 Dor in the LMC. At variance, a starburst galaxy displays violent star formation on such a scale that the whole system is out of balance with respect to available ressources. Perhaps the most fundamental definition of a starburst would be a galaxy in which the star-formation rate approaches the upper limit set by causality (Heckman, 2005, his Fig. 1). For a given total  mass and gas mass fraction the upper bound of the star formation rate (SFR) is proportionnal to $\sigma  {^3}$ where $\sigma$ is the stellar velocity dispersion (Murray et al. 2005). Other physical quantities are often considered such as the duration at which the current SFR consumes the remaining interstellar gas (the inverse of this is called the efficiency). In such context, a starburst produces the current stellar mass at the observed star formation rate on a time much shorter than a Hubble time. As pointed out by Kennicutt (1998) a starburst must develop a large SFR per unit area. Extreme starbursts have star formation rates per unit area larger than  $1M_{\odot}$ $yr^{-1}$ $ kpc^{-2}$.  These figures correspond to L(bol) $\geq 10^{10} L_{\odot}$ $kpc^{-2}$ hence thousands of times larger than normal discs galaxies and consumption times of only $10{^8}$ years.
It is important to realize that this definition refers to major starbursts and is somewhat arbitrary:  
starburst properties appear mostly continuous across the range of amplitudes observed.

\section{The Local Star-Forming Galaxies}
  Local Star-Forming galaxies provide rougly 10\% 
of the total radiated energy and roughly 20\% of all the high mass star formation 
in the present day (Brinchmann et al. 2004).
Among local star forming galaxies, sometimes referred as HII galaxies, 
most are dwarfs: dwarf irregulars 
(dIrr) and  Blue Compact Dwarfs (BCD). They remain, however, a minority among 
the general  population of dwarf galaxies (Kunth \&  {\" O}stlin, 2000; hereafter KO2000).
In term of absolute star  formation rates 
most figures are not very impressive. Dwarfs range from $M_{B}$ $\approx$ -14.0 to -17.0 and SFRs
 from $\approx$ 0.02 to $1M_{\odot} yr^{-1}$, while giants have $M_{B}$ up to -21
and reach SFRs of 20 to 40 $M_{\odot} yr^{-1}$ at most. These galaxies possess a gigantic
reservoir of HI gas that sometimes extends into huge halos 
surrounding the optical region.  The mechanism that ignites strong bursts of star
 formation across these systems is still not clear. Star formation may 
 be triggered during encounter events (merging or interacting galaxies)
  or stochastic processes within the galaxy itself.

At optical wavelengths, their spectra are dominated by young stars and
ionised gas, closely ressembling those of giant  HII regions in nearby spiral 
and irregular galaxies. Analysis of their spectra show that most of them 
are metal deficient. Towards the lower end of the metallicity distribution
(O/H $\sim 1/50 Z_\odot$) we find galaxies like IZw18 and SBS0335-052. 
Due to their extreme properties it has been conjectured by Searle \& Sargent (1970)
that these chemically unevolved galaxies could be young systems still in
the process of forming.

In star forming dwarfs, because of their low mass, star formation event occurs 
over a large part of the discs.  Some star forming regions are very compact and remain 
confined within the inner few hundreds pc of their hosts while others have one or several star forming 
clumps in off--nucleus regions.
  These events have strong and sometimes devastating impact on their different phases of the 
  interstellar medium. The hot gas, typically observed in X-rays, 
  can, at least in some cases, blow out from these galaxies (see Section 8). A starburst event is likely to
 provide an important mechanism for the metal enrichment of the intergalactic medium (IGM) in the early universe. 
 On small scales, some dwarf galaxies show numerous HI holes, 
   suggesting propagating star formation and allowing the investigation of early phases 
   of what later may become a blow-out of the gas. Molecular gas, the site of actual star formation, is notoriously difficult 
   to detect in dwarfs. The reasons for this are not necessarely related to a dearth of
    molecular material, but rather the low metallicity environment characteristic for dwarf galaxies and the low HI density that inhibit the rate of formation of the diffuse molecular gas. It is also likely that
     the diffuse molecular hydrogen is destroyed by the incident UV flux from the massive
      stars in the star-forming
region. Most of the remaining molecules should be in dense clumps, which are
opaque to far-UV radiation, and do not contribute to the observed spectra (Vidal-Madjar et al. 2000; Hoopes et al.
2004).

\section{The more distant starbursts}
At $ 0.4 \leq $ z $ \leq 1.0 $ Luminous Compact Blue Galaxies (LCBGs)  dominate the number 
density of galaxies at intermediate redshifts. They are small starburst systems of 
($R_{e}\leq 3.0$ kpc) that have evolved more than any other galaxy class in the last 8 Gyrs (Philipps et al. 1997, Guzman et al. 2003). They are a major contributor to the observed enhancement 
of the UV luminosity density of the universe at $z<1$, their number density decline being in concert
 with the rapid drop in the global Star Formation Rate since $z=1$. They appear to form a bridge in 
 redshift, size, and luminosity between Lyman-break galaxies and local H\textsc{II} galaxies today.
  On the other hand, the work of Steidel and  collaborators (e.g. Steidel et al. 1996)
has confirmed  a substantial population of star-forming galaxies at ~$z \sim 3$, with a  
comoving number density of roughly 10 to 50 \% that of present day luminous galaxies ($L \ge L^{\star}$). 
 It is clear that observational biases play a role since at $z\geq 2$,  L $\leq L{^*}$ 
  are more difficult to study.
 
%Let's remark that to define the metallicity of a galaxy requires some caution

\section{Measuring heavy element abundances from ionized gas}
 The spectral analysis of the HII galaxies
 shows that many HII galaxies are metal poor objects (probably the reason 
 I was invited to speak at this conference!), some of
them -- IZw18 and SBS\,0335-052  -- being among the most metal poor systems known.
Heavy element abundances are relatively easy to measure in star-forming  galaxies
because they contain ionized gas clouds in which large numbers of hot stars are
embedded. What is observed in the optical are narrow emission lines superimposed on a
blue stellar continuum. These are identified as helium and hydrogen
recombination lines and several forbidden lines: O, N, S, Ne, Ar, H and He lines
are currently measured (Izotov et al., 2006).
Methods used in determining abundances are well understood and generally more reliable
than those based on stellar absorption line data because transfer problems
become less important.
 
  Oxygen is the most reliably determined element, since the major
 ionisation stages can all be  observed. Moreover the [O{\sc iii}]${ \lambda  4363}$
  line allows an accurate 
determination of the electron temperature. The intrinsic uncertainty 
in this method 
is of the order of $\sim$0.1 dex. Furthermore, when the electron temperature 
cannot be determined, empirical relations
  between the 
oxygen abundance and the 
[O{\sc ii}]${\lambda 3727}$ and [O{\sc iii}]${\lambda\lambda 4959,5007}$ strength relative
to H$\beta$\ 
are used,
 though with lower accuracy (0.2 dex or worse), (Pagel, 1997). 
For other species, in general, all the ionisation stages are not seen 
 and an ionisation correction factor  
must be applied to derive their total abundances.

The ultraviolet (UV) region is dominated by the hot stellar continuum and shows relatively 
weak emission lines except for those that originate in stellar winds and has provided ways of measuring 
%However 
%there are a few notable exceptions and
%owing to the International Ultraviolet Explorer (IUE) and the Hubble 
%Space Telescope (HST) 
nebular carbon and silicon abundances.

We caution that many parameters control the observed metallicity in  a given galaxy. 
They are incorporated into chemical evolutionnary models and 
include : stellar evolution and nucleosynthesis, inflows and outflows
and the problem of the mixing and dispersion time scale of freshly released 
heavy elements (KO2000). Moreover, metallicity measurements may be relevant to only one 
particular component of a galaxy. A suggestion of  Kunth \& Sargent (1986)
has been made that HII gas could enrich itself with metals expelled by CC-SNe of time 
scales shorter than the lifetime of a starburst (self-pollution). Recent FUSE 
observations of some dwarf galaxies
show a possible disconnect in metallicity between HI and HII regions,
although  possible saturation effects on the line of sight 
may alter such a comparison (Lebouteiller et al. 2006).

One important aspect of H{\sc ii} regions abundances is that they can be obtained 
also at great distances. This makes them powerful tools also for studying 
high redshift galaxies, with the price that our view will be biased towards
actively star-forming systems. For a discussion on possible problems associated
with deriving abundances in very distant galaxies, see Kobulnicky et al. (1999).

\subsection{Do we expect metal-rich systems?}

 Knowledge of the chemical composition could be used to
constrain the age of any given galaxy provided that the heavy element abundance of
galaxies increases forever with age, and that the gas-phase
metallicity can be assumed to be equal to the metal abundance of the
stellar population. The first assumption assumes that galaxies evolve in a
``closed box'' manner. In this model, the metal content of a galaxy
should be a function of its gas mass fraction only. The predicting
power of this model would therefore be enormous, should this hypothesis be true for
at least some fraction of galaxies (but see Section 7). One can easily show - under the 
hypothesis of a Salpeter initial mass function and normal stellar yields that a natural limit would be reached
of the order of $\approx2Z_{\odot}$. To go  beyond this one would need to strongly bias the IMF 
towards high masses or/and use up all the gas (100\% efficiency!).
However, galaxies (and star-forming ones in particular) are not known to 
 deviate significantly from normal IMF while they are 
known to exchange large amounts of gas with the intergalactic medium 
(outflows, infall, merging etc..). Hence it can be understood that the requirements
applying to the cosmic enrichent of the Universe as a whole are not always 
 met for individual galaxies.  
 One can of course argue that observationnaly very
metal rich galaxies might escape from emission line surveys as their
 usual tracers such as the O{\sc iii}]${\lambda\lambda 4959,5007}$, for instance, become
extremely faint beyond 12+log(O/H)= 9.0 (Stasinska, 2002).

The second requirement is more complex to verify. First of all the
gas-phase metallicity does not necessarely directly relate to the heavy-element abundance
of a galaxy.  Second, it is very important
to keep in mind that metals ejected from dying stars do not necessarely mix instantly
into the general interstellar medium of galaxies (Roy \& Kunth, 1995).
This delay can be as large as one billion years (Tenorio-Tagle, 1996). 
This point is further discussed in Section 8.

\section{The Luminosity-Metallicity relation}

Star-forming galaxies will likely produce superwinds that will export metal-enriched
gas to the intergalactic medium. Merger events will also disrupt the 
``closed-box'' model paradigm. The strongest evidence from a departure from simple
close box models comes from the  luminosity-metallicity relation.
%The metal content of galaxies is an important diagnostic of galactic evolution. 
The luminosity--metallicity relationship, in such a context, is a useful relationship between
 evolutionary state and metallicity hence the location of galaxies in the $M_{B}$--$12+\log\mbox{(O/H)}$ plane 
is  of crucial importance to understand the details of the inner workings of starbursts. 
One reason for the existence of this relationship - assuming that more luminous galaxies 
are also more massive - is that more massive objects can lock up the metals created in their 
stars more easily than less massive galaxies because of their deeper 
gravitational wells.
A metallicity luminosity relation can also arise if smaller galaxies have larger gas fractions 
than larger galaxies. This is seen statistically in the local universe and can result from the fact that  smaller galaxies evolve more slowly 
(lower SFR per unit gas mass) than larger ones. 
In Fig. 1, we show the 
oxygen abundance vs. luminosity diagram for dIs (crosses), BCGs (filled symbols) and 
LSBGs (open symbols), based on data collected from the literature, with available 
abundances and integrated B magnitudes. At first sight many BCGs do not appear extreme 
when compared to the normal dIs. Indeed some BCGs are more metal rich than dIs at a 
given luminosity, while the opposite would be expected if BCGs were bursting dIs. 
These BCGs  may be in a post burst stage and the fresh metals may have 
become ``visible'' already. Secondly, some  ``extreme
BCGs'' appear  much more metal-poor at a given luminosity. These extreme BCGs (XBCGs as listed in KO2000) 
are 3 magnitudes brighter or more at a given metallicity, or equivalently 0.5 dex less 
metal rich at a given luminosity  as compared to the dI relation. The intriguing XBCGs include 
IZw18 and SBS~0335-052, Tololo~1924-516) (see \"Ostlin et al. 1999, 2001)  
among others. 

\begin{figure}
\includegraphics[width=5.0in,height=5.0in] {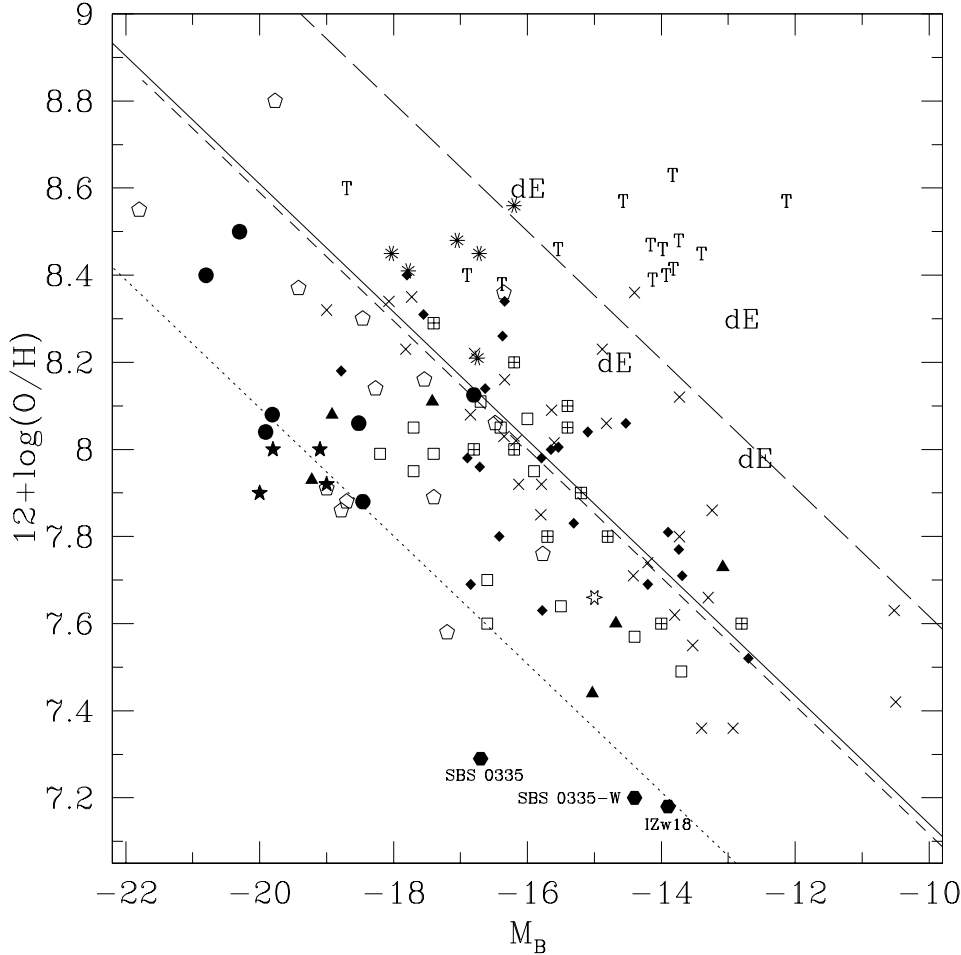}
\hfill
%\vspace{-2cm}
%\hspace{1cm}
%\resizebox{12cm}{!}{\includegraphics{metlum2_ok.ps}}
%\vspace{-0.5cm}
\caption{{\small The luminosity versus metallicity diagram for dwarf galaxies. 
Crosses represent dIs.
% taken from Richer and McCall (1995) and Skillman et al. (1989).
Filled symbols represent galaxies classified as BCGs or H{\sc ii}-galaxies: Small filled
diamonds are ``regular'' galaxies which  can be classified as Type II or iE/nE, 
while  filled circles are 
galaxies that can be classified as Type I.
%(see text for description of types). 
 Filled triangles are BCGs
for which no classification or images were available. The three most metal-poor galaxies are
labelled and shown as filled hexagons. Filled 
stars are luminous BCGs.
%from \"Ostlin et al. (1999a,b), three of which are of Type I.  
Asterisks show the location of ``blue amorphous galaxies'' 
%(Marlowe et al. 1999,
 except 
for II~Zw40 which is the filled circle falling on the short dashed line). 
LSBGs are shown as open squares 
%(blue LSBGs, Bergvall et al. 1999, R\"onnback and Bergvall 1985)
 and open pentagons 
 %(McGaugh 1994, McGaugh and Bothun 1994).
Open star is H{\sc i}~1225+01, the H{\sc i}-cloud in Virgo.
%(Salzer et al. 1991). 
Boxes with plusses 
inside are quiescent (dI/LSBG) dwarfs.
 %from van Zee et al. (1997b,c). 
 Candidate tidal dwarfs 
 %(Mirabel et al. 1992, Duc and Mirabel 1994, 1998)
 are shown as ``T'', and dEs  are shown as ``dE''.
 %(data from Mateo 1998). 
The solid line shows the $M_B - O/H$
relation for dIs from Richer and McCall (1995), while the dotted line shows the same relation
offset by 3.5 magnitudes, indicating the location of XBCGs. The short dashed line shows the 
$M_V - Z$ relation for dE/dSph from
Caldwell (1998) assuming $(B-V) = 0.75$ and [O/H] = [Fe/H], while the long dashed line shows
the same relation assuming  [O/H]  = [Fe/H] + 0.5.
%When necessary, we have rescaled absolute magnitudes to H$_0=75$km/s/Mpc.
}
%{\scriptsize Metallicities of BCGs  from:
%Izotov and Thuan (1999), Lipovetsky et al. (1999), Bergvall and \"Ostlin (1999), Telles and
%Terlevich (1997), van Zee et al. (1998), Kunth and Joubert (1985), Alloin et al. (1978),
%Thuan et al. (1996), Masegosa et al. (1994). 
%Absolute B-magnitudes for BCGs from: Telles and Terlevich (1997, assuming $B-V=0.5$), 
%Papaderos et al. (1996a, 1998), \"Ostlin et al. (1999a), Thuan et al. (1996), Mazzarella and 
%Boroson (1993), Salzer et al. (1989a), Schulte-Ladbeck et al. (1998). 
Details and complete references for the object sources, their B magnitudes and metallicities 
are listed in KO2000.}

\end{figure}
%
%
%
%
%While enriched galactic winds  
%could possibly explain the extreme metal deficiency of the least 
%massive dwarfs, the existence of XBCGs in general cannot be understood in this way (infall of fresh gas 
%could also be an alternative). 
%It is therefore very important to understand the position of a galaxy in the $M_{B}$--$12+\log \mbox{(O/H)}$ plane. 
%Some luminous dwarf galaxies turn to be metal--poor, it would be interesting to know
%how these objects happen to attain such large luminosities, given that
%the number of different stellar generations should be small. If, they are metal-rich objects for their luminosity, 
%the question would then be: how do such small systems retain the metals they have created?
Tidal dwarfs appear metal--rich for their metallicity because they were formed from the already enriched gas of their parent galaxies (Duc \& Mirabel, 1998).
Some arguments 
have been  given that ascribe the luminosity-metallicity correlation to the effects of selective losses of heavy elements from galaxies in supernova-driven outflows. A reduction in the effective yields
(most likely due to galactic winds) can explain the observations: Garnett 
(2002) has shown that metal loss due to supernova-driven 
 winds is a process at work in Irr and some spiral galaxies. Indeed, galaxies with low
  rotational velocities (hence low mass) loose a large fraction of metals while 
  galaxies above this value tend to retain metals. Not only is the 
metallicity-luminosity relation of galaxies inferred from the SDSS  providing the best 
evidence for galactic winds (Tremonti et al. 2004) but this relationship is 
expected to evolve with cosmic epoch. 

At large redshifts ($z \geq 2.0$), and using emission-line ratios, 
Pettini et al. (2001, and references therein)
already found evidence that Lyman Break Galaxies (LBGs) of 
under-solar abundance are 5--40 times more luminous than local systems of similar metal 
content, confirming that the luminosity-to-metal ratio varies with time.

\section{Cosmological context}

It has been possible to 
sketch the star-forming history 
of the Universe at high redshifts (Chary \& Elbaz, 2001). It appears 
that the overall SFR of the galaxy population seems to 
increase from ~$z=0$~ to ~$z=1$;  this epoch is believed to be that of disc formation.
 The cosmic star formation rate becomes constant or slightly declining from $z= 1$~ 
 to ~$z= 4$~ and this period 
is sometimes refered to as the spheroid epoch.
In fact, at $z\geq 1$ more galaxies were in interactions, 
starbursts built on rapid time scales with  high 
SFR/unit area  hence metallicity could be high. At high redshift, the highest 
SFRs occur 
in the most dusty galaxies (sub-mm sources) telling us that a young galaxy does not 
necessarely mean being metal poor! 
 Numerical simulations indicate that metal production 
occurs at early times (due to spheroids consecutive to the  strong starburst they experience)
 while at  $z \leq 2$  the main sites of element production 
 become the spiral discs (Calura \& Matteucci 2004). Local irregular systems have a negligible 
 contribution to the  total element production. \\
Numerical simulations by Cen and Ostriker (1999) predict the evolution of the 
metal content of the Universe as a function of density, by incorporating star 
formation and its feedback on the IGM. At a given gas density (corresponding e.g. 
to a rich cluster, a disc galaxy, dwarf etc...) their models predict an evolution 
with redshift, but more importantly
show that metallicity is  a stronger
function of density than age; moreover with a considerable scatter. 
At low redshift, one would expect a few percent of the gas-rich dwarfs to 
have metallicity on the order of IZw18, without invoking their youth.
This  is also suggested  by the  Ly$\alpha$ absorption 
systems, (Shull et al. 1998).
 Ferrara and 
Tolstoy (2000) discussing the feedback from star formation 
in dwarf galaxies argue from this that low mass galaxies could 
be a major contributor to  metals in the IGM.

%The idea of so called  population III objects has been discussed for quite a while.
%These hypothetical objects would be (or host) the first generation of stars, appearing 
%before the main epoch of galaxy formation, and thus be a source of pre enrichment in the 
%Universe. The subject remains speculative, but the existence of Pop III stars
%cannot be ruled out (Cayrel 1996).

\section{Fate of metals}

\subsection{The mixing time scales}

The fraction of metals that mixes within the ISM crucially depends on the
 thermodynamical status of the ISM and on the star formation history 
 of the galaxy.  The very first burst of star formation must occur in a 
 very cold and dense medium hence radiative losses of the expanding bubbles and
superbubbles become significant (i.e. comparable to the thermal energy
of the cavity) in a short time-scale of the
order of 10 Myr (MacLow \& McCray, 1988; Recchi et al. 2001). 
At later times, the ISM  being warmer, more diluted 
and irregularly structured, radiative losses  will be
smaller, therefore the impact of late generation of stars into the ISM
will be stronger. 
Freshly produced metals by these late generations
of stars, will be either directly channeled along the galactic outflow
or  released in a warm and tenuous medium on a much larger mixing time-scales - 
as large as one billion year (Recchi et al. 2004; Tenorio-Tagle, 1996).
%This scenario clearly contradicts the self enrichment
%picture within the ionised gas - unless an important fraction of hot gas has rapidly cooled down -. 

In more general terms, 
Roy \& Kunth (1995) conceede 
that dwarf galaxies are expected to show kpc scale abundance inhomogeneities.
From an observational point of view, some galaxies support the fast mixing
scenario  while some do not. This is true in NGC~5253, IIZw~40 and Mrk 996 where local N/H
 overabundances has been attributed to localised pollution from WR stars
  (Kobulnicky et al. 1997; Walsh \& Roy, 1993; Thuan et al. 1996). 
  This indicates that rapid mixing can take place. However this is not true for 
all young starbursts even when WR stars are suspected or well observed
(Oey \& Shields, 2000). On the other hand
IZw~18 appears to be rather homogeneous (Legrand et al. 2000)
hence does not advocate in favor of the concept of rapid self-enrichment. 
Arguments in favor of complete mixing over long time scales lie from the observation that 
disconnected H\,{\sc ii} regions within the same galaxies have nearly the same
abundances. Six H\,{\sc ii} regions in the SMC have log O/H = 8.13 ($\pm 0.08$)
while 4 in the LMC give log~O/H = 8.37 ($\pm 0.25$) (Russell \& Dopita 1990).

The possibility that  metallicities in the neutral gas phase are 
orders of magnitude below the H\,{\sc ii} region abundances  would be
an ultimate test of large scale inhomogeneities. Kunth and Sargent (1986) 
proposed a clean test for the self-enrichment hypothesis by using a QSO as background source 
to measure the abundance of metals in
the neutral gas. Such a test was recently performed by Bowen et al (2005) who measured 
the abundance of sulfur in the neutral gas of the dwarf SBS 1543+ 593 to be the same  as that 
of oxygen in one of its neighbouring HII region - at variance with self-enrichment pollution
expectations.   
A good benchmark for the enrichment process in the early evolution of a
galaxy is our Milky Way.  Metal-poor halo stars (Cayrel et
al. 2004) show a small spread in  alpha elements.
This
%conflicts with the inhomogeneous enrichment process adopted by
%Argast et al. (2000) but 
is in agreement with the
instantaneous mixing hypothesis commonly assumed in simple chemical
evolution models (e.g. Francois et al. 2004).  Arnone et al. (2005)
 confirmed the
need for a fast mixing process for the alpha elements, although the
large scatter in [Ba/Fe] shows that ejecta of stars in different mass
ranges could have different mixing timescales.

\subsection{The IGM enrichment}

It is an important question to decide whether the newly produced metals leave the ISM and 
pollute the IGM.  In principle, the energy released by stellar winds at the very early stage of a starburst
 (Leitherer et al. 1992) and supernovae explosions after a
burst of star formation often exceeds the binding energy of dwarf
galaxy, therefore the development of a galactic wind is likely...but:

{\bf Observations}: Nearby Far-Infrared galaxies support the occurrence of
galactic winds. The nearest edge-on disk galaxies show clear kinematic
signatures of an outflow along the minor axis with velocities of the
order of 200 to 600 km sec$^{-1}$ (Heckman, Armus \& Miley, 1990).
As shown by HST, FUSE and Chandra observations, clear signatures of an outflow are
 present in many  other nearby galaxies such as NGC1705 (Heckman et al. 2001), 
 NGC1569 (Martin et al. 2002), NGC3079 (Cecil et al. 2001), 
while that molecular disk-halo outflow is occurring in NGC3628 (Irwin \&
Sofue, 1996). Kunth and collegues (see Mas-Hesse et al. 2003)
 have shown that the escape of 
the Ly$\alpha$ photons in star-forming galaxies  strongly depends on the 
dynamical properties of  
their interstellar medium. The Lyman alpha profile in Haro~2
indicates a superwind of at least 
200 km/s, carrying a mass of $\sim 10^7 M_{\odot}$, which can
be independently traced from the H$\alpha$ component (Legrand et al. 1997).
However,  high speed winds do not necessary carry a lot of mass.
Martin (1996)  argues that a bubble seen in IZw~18 
 will ultimately blow-out together with its hot gas component.  
Diffuse X-ray emissions around starburst galaxies correlating with the
star formation per unit area (Strickland et al. 2004),
corroborate the idea that these are the results of outflows driven
by the starburst activity.
The best examples of large-scale outflows driven by SNe feedback are
however perhaps found  at large redshifts (Pettini et al.  2001).

{\bf Interpretations}: Even in spite
of the ubiquity of outflow phenomena, there is no certainty that the
outflowing gas will not recollapse towards the center of a galaxy in
the future. 
%Since little in known about the details of the interactions between the evolving 
%supernova remnants, massive stellar bubbles and the ISM
 It is  possible that 
an outflow takes the fresh metals with it and in some cases leaves a galaxy 
totally cleaned of gas but the hot gas outside of the H\,{\sc ii} regions may simply stay 
around in the halo. The presence of outflows has been used as an indication that supernova  products 
and the whole of the interstellar medium is easily ejected from the host dwarf systems,  
causing the contamination of the intra-cluster medium (Dekel \& Silk 1986; 
De Young \& Heckman 1994). This type of assumption is currently blindly used by
cosmologists in their model calculations.

In a more refined approach MacLow \& Ferrara (1999) models with $\sim 10^6 M_{\odot}$  systems experience a complete blow-away, 
whereas in more massive ones only a small fraction of the gas mass escapes the galactic potential
well.  In these blow-out phenomena a very large fraction of
metals is lost and for models with masses below or equal $\sim 10^8M_{\odot}$
 almost all the metals are lost through the galactic wind.  This
result (that metals are ejected much more easily that pristine ISM) is
pretty common (the same found Strickland \& Stevens 2000 or D'Ercole \&
Brighenti 1999) but is at variance with Silich \&
Tenorio-Tagle vision (2001). They argue that the indisputable presence of metals 
in galaxies implies that  
supernova products are not completely lost in all 
cases: in a dwarf galaxy which has a weaker gravitational potential, these effects 
may result in gas loss from the galaxy {\bf unless} the presence
of a low HI density halo acts as a barrier. 
Model  calculations developed by Silich \& Tenorio-Tagle (1998)
 predict that  superbubbles in   amorphous  dwarf galaxies must have already 
undergone blowout and are presently
evolving into an extended low-density halo. This should inhibit the loss 
of the swept-up and processed matter into the IGM. Recent Chandra X-rays  
observations are mitigated: some young starbursts indicate metal losses 
 from  disks (see Martin, Kobulnicky \& Heckman 2002) but not in  the 
case of NGC~4449 with an extended H\,{\sc i} halo of around 40~kpc 
(Summers et al. 2003). 
%region of $\sim 100$ pc size. During the supernova phase the continuous energy 
%input rate from coeval starbursts or continuous star--forming episodes, 
%maintains the temperature of the ejected matter  
%above the recombination limit ($T \sim 10^6$ K) allowing superbubbles to 
%reach dimensions in excess of 1 kpc. 
%he interstellar medium gas and generates gas flows. The 
%properties and evolution of these flows ultimately determine the fate of the 
%newly formed metals and the manner they  mix with the original interstellar 
%medium. 

Legrand et al (2001) have compared Mac Low \& Ferrara (1999) and Silich \& Tenorio-Tagle (2001) 
theoretical estimates with some well--studied starburst galaxies.  
Values of the derived mechanical energy injection rate were compared with their
 hydrodynamical models predictions. 
The net result  is that all galaxies  lie  
above the lower limit first derived by  Mac Low \& Ferrara (1999)
for the ejection of metals out of flattened  disk-like ISM density 
distributions energized while most are {\bf below} the limit for the 
low density halo picture.

\subsection{Which galaxies enrich?}

Which galaxies experience galactic winds and contribute to the chemical 
enrichment of the IGM? Gibson \& Matteucci (1997) showed that dwarf
galaxies can provide at most 15\% of the ICM (intra cluster medium), giant ellipticals
being responsible for 20\% of it and the remaining 65\% being
primordial.   
 It has
been  proposed that only certain elements are lost (or in different
proportions) hence reducing the effective net yield of those metals as 
compared to a simple chemical evolution model (Edmunds, 1990). 
The SNe involved in such a wind are likely to be of type II because type Ia 
SNe explode in isolation and will less likely trigger chimneys from which
metals can be ejected out of the plane of a galaxy. In this framework O and
 part of Fe  are lost while He and N 
(largely produced by intermediate stars) are not. This would result in a 
cosmic dispersion in element ratios such as N/O between galaxies that 
have experienced mass loss and those that have not. 
In general, as we discussed above, the flatter a galaxy is, the easier is the development of
a galactic wind (Strickland \& Stevens, 2000),
therefore for elliptical galaxies the development of a galactic wind
occurs only when the total thermal energy overcomes the binding energy
of the galaxy, condition not required in flattened systems.  However, 
the initial star formation rate in elliptical galaxy should be large
enough to fulfill the condition for the onset of a galactic wind.
Therefore IGM pollution due to galactic winds from giant ellipticals
is likely.

Dwarf spheroidal galaxies with a more shallower potential well 
should favor the
development of a galactic wind and their lack of gas
might be explained by this phenomenon.  But  this is not as easy at it looks,  
Marcolini, D'Ercole \&
Brighenti (2006) simulating a spherically symmetric galaxy 
such as Draco 
failed in producing a galactic wind hence the problem is difficult to
solve from the hydrodynamics and I invite the reader to look at the
paper of Skillman \& Bender (1995) for a thorough discussion of these problems.

A very last point concerning iron was brought to my attention by S. Recchi.
  A very large amount of iron is
in the ICM, perhaps exceeding the amount of alpha elements (e.g. M87;
Gastaldello \& Molendi 2002).  Since 
galactic winds should be enriched in alpha elements as they
are released by more numerous SNeII there
should be an overabundance of alpha-elements, at odds with
observations.  However, elements produced by SNeIa are
easily channeled along the wind (the hole in the ISM is already
there, releasing the efforts to leave the parent galaxy),
therefore  simulations of starburst galaxies give larger
ejection efficiencies for iron-peak elements. 
Models with
continuous star formation have similar ejection efficiencies of alpha- and
iron-peak elements.

%In general, after a bursting star formation the ejection
%efficiency of N is larger than the ejection efficiency of O.  This is¨
%because N is produced in timescales similar to the SNeIa timescales,
%therefore it can be pushed out of the galaxy by the energy released by
%these SNe. Their thermalization efficiencies are in fact larger than
%for SNeII because they explode in a warmer and more tenuous ISM (they
%begin to explode after all the SNeII have already exploded).  This
%difference is negligible for models with continuous star formation
%because in this case, after the first few tens of Myr, both SNeII and
%SNeIa release metals in the same ISM (Recchi).

\subsection{Acknowlegments:} I thank F. Calura, D. Garnett, C. Hoyos, M. Mas-Hesse, G. \"Ostlin, S. Recchi and I. Stevens for stimulating discussions.


\begin{thebibliography}{99}


%\bibitem[]{ar2000} Argast, D., Samland, M., Gerhard, O.~E., \& Thielemann, F.-K.\ 2000, A\&A, 356, 873 

\bibitem[]{ar2005} Arnone, E., Ryan, S.~G.,  Argast, D., Norris, J.~E., \& Beers, T.~C.\ 2005, A\&A, 430, 507 


%\bibitem[]{Barbieri&Gratton2002} Aloisi, A., Tosi, M., \& Greggio, L.\ 1999, \aj, 118, 302 

%\bibitem[]{Barbieri&Gratton2002} Aloisi, A., Tosi, M., \& Greggio, L.\ 2001, \apss, 276, 421 

%\bibitem[]{Barbieri&Gratton2002} Aloisi, A., Savaglio, S., Heckman, T.~M., Hoopes, C.~G., Leitherer, C., \& Sembach, K.~R.\ 2003, AAS meeting, 202, 1403  

%\bibitem[]{Barbieri&Gratton2002} Dekel, A. \& Silk, J., 1986, \apj, 303, 39 


\bibitem[]{bo2005} Bowen, D.~V., Jenkins, E.~B., Pettini, M., \& Tripp, T.~M.\ 2005, ApJ, 635, 880 

\bibitem [ ] {} Brinchmann J., et al. 2004, MNRAS,  351, 1151


\bibitem[]{ca2004} Calura, F., \&  Matteucci, F.\ 2004, MNRAS, 350, 351 

%\bibitem[]{ca2005} Cannon, J.~M., Skillman, E.~D., Sembach, K.~R., \& Bomans, D.~J.\ 2005, ApJ, 618, 247 

\bibitem[]{cay2004} Cayrel, R., et al.\ 2004, A\&A, 416, 1117 

\bibitem[]{ce2001} Cecil, G., Bland-Hawthorn, J., Veilleux, S., \& Filippenko, A.~V.\ 2001, ApJ, 555, 338 

\bibitem[]{ce1999} Cen R., Ostriker J.P. 1999, ApJ, 519, L109

\bibitem[]{ch2001} Chary, R., \& Elbaz, D.\ 2001, ApJ, 556, 562 

\bibitem[]{de1986} Dekel, A., \& Silk, J.\  1986, ApJ, 303, 39 

\bibitem [] {} De Young, D. \& Heckman, T., 1994, ApJ, 431, 598 

\bibitem []{de1999} D'Ercole, A., \& Brighenti, F.\ 1999, MNRAS, 309, 941 

\bibitem[]{du1998} Duc, P.-A., \& Mirabel, I.~F.\ 1998, A\&A, 333, 813 

%\bibitem [ ] {}  Dennerl, K.~et al.\  2001, \aap, 365, L202  

\bibitem [ ] {} Edmunds, M.~G.\ 1990, MNRAS, 246, 678 

\bibitem[]{fe2000} Ferrara, A., \& Tolstoy, E.\ 2000, MNRAS, 313, 291 

\bibitem[]{fr2004} Fran{\c c}ois, P. et al. 2004, A\&A, 421, 613 

\bibitem[]{ga2002} Garnett, D.~R.\ 2002, ApJ, 
581, 1019 

\bibitem[]{ga2002} Gastaldello, F., \& Molendi, S.\ 2002, ApJ, 572, 160 

\bibitem[]{gi1997} Gibson, B.~K., \& Matteucci, F.\ 1997, ApJ, 475, 47 

%\bibitem[]{guz97} Guzm\'an R., Gallego J., Koo D.~C., Phillips A.~C., Lowenthal J.~D., Faber S.~M., Illingworth G.~D., Vogt N.~P., 1997, ApJ, 489, 559 

%\bibitem[]{guz98} Guzm\'an R., Jangren A., Koo D.~C., Bershady M.~A., Simard L., 1998, ApJ, 495, L13 
 
\bibitem[]{guz03} Guzm{\'a}n R., {\"O}stlin G., Kunth D. et al. 2003, ApJ, 586, L45 
 

\bibitem[]{ha2002} Hamann, F., Korista, K.~T., Ferland, G.~J., Warner, C., \& Baldwin, J.\ 2002, ApJ, 564, 592 

%\bibitem [ ] {} Heckman T. 1994, in: Mass-Transfer Induced Activity in Galaxies, Shlossman I., ed., (CUP: Cambridge) p. 234

\bibitem[]{he1990} Heckman, T.~M., Armus, L., \& Miley, G.~K.\ 1990, ApjS, 74, 833 

\bibitem[]{he2001} Heckman, T.~M. et al. 2001, ApJ, 554, 1021 

\bibitem [] {} Heckman T. 2005, ASSL: Vol 329; Starbursts: From 30 Dor to Lyman Break Galaxies, p. 3 


\bibitem[2004]{Hoopes et al.2004} Hoopes C.~G., Sembach K.~R., Heckman
T.~M. et al.\ 2004, ApJ, 612, 825

\bibitem [ ] {} Hoyos C., 2006, On the Nature of Distant LCBs; Thesis Un. Autonoma de Madrid

%\bibitem [ ] {} Hensler G., Rieschick A., 1998, in Highlights in Astronomy, Ed. Andersen J., Vol. 11A, p 139, (Kluwer)

%\bibitem [ ] {} Hunt, L.~K., Thuan, T.~X., \& Izotov, Y.~I.\ 2003, \apj, 588, 281 

\bibitem[]{ir1996} Irwin, J.~A., \& Sofue, Y.\ 1996, ApJ, 464, 738 


%\bibitem [ ] {} Izotov,  Y.~I., Thuan, T.~X., \& Lipovetsky, 
%V.~A.\ 1997, \apjs, 108, 1

%\bibitem [ ] {} Izotov, Y.~I.~\&  Thuan, T.~X.\ 1999, \apj, 511, 639  

%\bibitem [ ] {} Izotov, Y.~I., Chaffee, F.~H., Foltz, C.~B., Green, R.~F., Guseva, N.~G., \& Thuan, T.~X.\ 1999, \apj, 527, 757 

%\bibitem [ ] {} Izotov, Y.~I., Schaerer, D., \& Charbonnel, C.\ 2001, \apj, 549, 878 


\bibitem[]{} Izotov, Y.~I., Stasi{\'n}ska, G., Meynet, G., Guseva, N.~G., \& Thuan, T.~X. 2006, ApJ, 448, 955 
			
     

\bibitem [ ] {} Kennicutt R. 1998, ApJ, 498, 541

%\bibitem [ ] {} Kobulnicky H.A., 1998, in ``Abundance Profiles: Diagnostic Tools for Galaxy History'', ASP Conf. Ser. Vol. 147, p. 108

\bibitem [ ] {} Kobulnicky H.A., Skillman E.D., Roy J.-R., Walsh J.R., Rosa M.R., 1997, ApJ, 477, 679 

\bibitem[]{ko1999} Kobulnicky, H.~A., 
Kennicutt, R.~C., Jr., \& Pizagno, J.~L.\ 1999, ApJ, 514, 544

%\bibitem[]{kob_koo_2000} Kobulnicky H.~A., Koo D.~C., 2000, ApJ, 545, 712 

\bibitem [ ] {} Kunth D.~\& Sargent W.~L.~W.\ 1986, ApJ, 300, 496 

%\bibitem [ ] {} Kunth D., Lequeux J., Sargent W.L.W., Viallefond F., 1994, \aap, 282, 709

%\bibitem [ ] {} Kunth, D., Mas-Hesse, J.M., Terlevich, E., Terlevich, R., Lequeux, J. \& Fall, S.M.\ 1998, \aap, 334, 11 

\bibitem [ ] {} Kunth, D. \& \" Ostlin, G. 2000, A\&ARev, 10, 1 : (KO2000)

%\bibitem [ ] {} Lebouteiller, V., 2003, \aap, submitted

\bibitem [ ] {} Lebouteiller, V., Kunth D. et al., 2006, A \& A, accepted

%\bibitem [ ] {} Lecavelier, A. et al., 2002, FUSE Sci. and Data Workshop, unpublished

%\bibitem [ ] {} Lecavelier, A. et al., 2003, \aap, submitted

%\bibitem [ ] {} Legrand, F.\ 2000, \aap, 354, 504 

\bibitem [ ] {}  Legrand, F., Kunth, D., Mas-Hesse, J.~M., \& Lequeux, J.\ 1997, A\&A, 326, 929 

\bibitem [ ] {}  Legrand, F., Kunth, D., Roy, J.-R., Mas-Hesse, J.~M., \& Walsh, J.~R.\ 2000, A\&A, 355, 891 

\bibitem [ ] {} Legrand F., Tenorio-Tagle G.,  Silich S., Kunth D., \& Cervi\~no M.\ 2001, ApJ, 560, 630

%\bibitem [ ] {} Lequeux, J.~\& Viallefond, F.\ 1980, \aap, 91, 269 

%\bibitem [ ] {} Lequeux, J., Kunth, D., Mas-Hesse, J. M., Sargent, W. L. W., 1995,  \aap, 301, 18 

\bibitem [ ] {} Leitherer, C., Robert, C., \& Drissen, L.\ 1992, ApJ, 401, 596 

\bibitem [ ] {} Mac Low, M.~\& Ferrara, A.\ 1999, ApJ, 513, 142 

\bibitem[]{mc1988} Mac Low, M.-M., \& McCray, R.\ 1988, ApJ, 324, 776 

%\bibitem [ ] {} Marlowe, A.~T., Heckman, T.~M., Wyse, R.~F.~G., \& Schommer, R.\ 1995, \apj, 438, 563 

\bibitem[]{ma2006} Marcolini, A., D'Ercole, A., Brighenti, F., \& Recchi, S.\ 2006, MNRAS, 371, 643 

\bibitem [ ] {}  Martin, C.~L.\ 1996, ApJ, 465, 680 

%\bibitem [ ] {} Martin, C.~L.\ 1998, \apj, 506, 222 

\bibitem [ ] {} Martin, C.~L., Kobulnicky, H.~A., \& Heckman, T.~M.\ 2002, ApJ, 574, 663 

\bibitem[]{ma2003} Mas-Hesse, J.~M., Kunth, D., Tenorio-Tagle, G., Leitherer, C., Terlevich, R.~J., \& 
Terlevich, E.\ 2003, ApJ, 598, 858 

%\bibitem [ ] {} Mas-Hesse, J.M., Kunth D., Tenorio-Tagle, G.,  Leitherer, C.,  Terlevich, E. \& Terlevich, R.\ 2003 december, \apj, in press

%\bibitem [ ] {} Matteucci, F.~\& Chiosi, C.\ 1983, \aap, 123, 121 

%\bibitem[Melnick et al.(1985)]{} Melnick, J., Terlevich, R., \& Eggleton, P.~P.\ 1985, \mnras, 216, 255 

%\bibitem [ ] {} Melnick J., Heydari-Malayeri M., Leisy P., 1992, \aap, 253, 16


\bibitem [ ] {}Murray N, Quataert E., Thompson T., 2005, ApJ, 618, 569 

\bibitem [ ] {} Oey, M.~S.~\& Shields, J.~C.\ 2000, ApJ, 539, 687 



\bibitem [ ] {} \"Ostlin G., Amram P., Masegosa J., Bergvall N., Boulesteix J., 1999, A\&AS 137, 419 


\bibitem [ ] {} \"Ostlin G., Amram P.,  Bergvall N., Masegosa J., Boulesteix J., Marquez I. 2001, A\&A 373, 800 

%\bibitem [ ] {} {\" O}stlin, G.\ 2000, \apjl, 535, L99 

\bibitem[]{pa1997} Pagel, B.~E.~J.\ 1997, Nucleosynthesis and Chemical Evolution of Galaxies, Cambridge Un. Press,  

%\bibitem [ ] {} Pantelaki I., Clayton D.D., 1987, in ``Starbursts and galaxy 
%evolution'', Eds. Thuan T.X., Montmerle T., Tran Thanh Van J., \'Editions Fronti\`eres, p. 145 

%\bibitem [ ] {} Petrosian, A.~R., Boulesteix, J., Comte, G., Kunth, D., \& Lecoarer, E.\ 1997, \aap, 318, 390 

%\bibitem [ ] {} Pettini M., Lipman K., 1995, \aap, 297, 63 


%\bibitem[]{pe1998} Pettini, M., Kellogg, M., Steidel, C.~C., Dickinson, M., Adelberger, K.~L., \& Giavalisco, M.\ 
%1998, ApJ, 508, 539 


\bibitem[]{pett01} Pettini M., Shapley A.~E., Steidel C.~C., Cuby J.-G., Dickinson M., Moorwood A.~F.~M., Adelberger K.~L., Giavalisco M., 2001, ApJ, 554, 981 
 

\bibitem[]{phi97} Phillips A.~C., Guzm\'an R., Gallego J., Koo D.~C., Lowenthal J.~D., Vogt N.~P., Faber S.~M., Illingworth G.~D., 1997, ApJ, 489, 543 

 
\bibitem[]{} Recchi, S., Matteucci, F., \& D'Ercole, A.\ 2001, MNRAS, 322, 800 

\bibitem[]{} Recchi, S., Matteucci, F., D'Ercole, A., \& Tosi, M.\ 2004, A\&A, 426, 37 
 
\bibitem [ ] {} Roy, J.-R. \& Kunth, D., 1995, A\&A, 294, 432

\bibitem [ ] {} Russell, S.~C.~\& Dopita, M.~A.\ 1990, ApJS, 74, 93 


\bibitem [ ] {} Sargent W.L.W., Searle L. 1970, ApJ, 162, L155 

\bibitem[]{sh1998} Shull, J.~M. et al.\ 1998, AJ, 116, 2094 

\bibitem[]{st2002} Stasinska, G.\ 2002, Revista 
Mexicana de Astronomia y Astrofisica Conference Series, 12, 62 

\bibitem [ ] {} Silich, S.~A.~\& Tenorio-Tagle, G.\ 1998, MNRAS, 299, 249 

\bibitem [ ] {} Silich, S.~\& Tenorio-Tagle, G.\ 2001, ApJ, 552, 91 

\bibitem[]{sk1995} Skillman, E.~D., \&  Bender, R.\ 1995, Rev Mex. de Astronomia y Astrofisica Conf. Ser., 3, 25 

%\bibitem [ ] {} Skillman E.D., Kennicutt R.C., 1993, \apj, 411, 655

%\bibitem [ ] {} Sofia, U.~J.~\&  Jenkins, E.~B.\ 1998, \apj, 499, 951  

 
\bibitem[]{2steidel96} Steidel C.~C., Giavalisco M., Pettini M., Dickinson M., Adelberger K.~L., 1996, ApJ, 462, L17 

\bibitem[]{st2000} Strickland, D.~K., \& Stevens, I.~R.\ 2000, MNRAS, 314, 511 

\bibitem[]{st2004} Strickland, D.~K., Heckman, T.~M., Colbert, E.~J.~M., Hoopes, C.~G., \& Weaver, K.~A.\ 2004, ApJ, 606, 829 

\bibitem [ ] {} Summers, L.~K., Stevens, I.~R., Strickland, D.~K., \& Heckman, T.~M.\ 2003, MNRAS, 342, 690 

\bibitem [ ] {} Tenorio-Tagle, G., 1996, AJ, 111, 1641

%\bibitem [ ] {} Tenorio-Tagle, G., Silich, S.A., Kunth, D., Terlevich, E. \& 
%Terlevich, R.\ 1999, \mnras, 309, 332 

\bibitem [ ] {} Thuan T.X., Izotov Y.I., Lipovetsky V.A., 1996, ApJ, 463, 120

%\bibitem [ ] {} Thuan, T.~X., Izotov, Y.~I., \& Lipovetsky, V.~A.\ 1997, \apj, 
%477, 661 

%\bibitem [ ] {} Thuan, T.~X., Lecavelier des Etangs, A.,  \& Izotov, Y.~I.\ 2002, 
%\apj, 565, 941  


\bibitem[Tremonti et al. (2004)]{tr2004} Tremonti, C.~A., et al. 2004, ApJ, 613, 898

%\bibitem [ ] {} van Zee, L., Westpfahl, D., Haynes, M.~P., \& Salzer, J.~J.\ 1998, \aj, 115, 1000 

%\bibitem [ ] {}  Viallefond F., Lequeux J., Comte G., 1987, in ``Starbursts and galaxy evolution'',  \'Editions Fronti\`eres, p. 139
  
\bibitem [ ] {} Vidal-Madjar, A.~et al.\ 2000, ApJ, 538, L77 


%\bibitem [ ] {} V\'ilchez J.M., Iglesias-P\'aramo J., 1998, \apj, 508, 248

\bibitem [ ] {}  Walsh J.R., Roy J.-R., 1993, MNRAS, 262, 27 
	 
\end{thebibliography}
\end{document}